\documentclass[conference]{IEEEtran}

\usepackage[T1]{fontenc}
\usepackage{amsmath,amssymb}
\usepackage{algorithm}
\usepackage{algpseudocode}
\usepackage{array}
\usepackage{booktabs}
\usepackage{cite}
\usepackage{float}
\usepackage{flushend}
\usepackage{graphicx}
\usepackage{microtype}
\usepackage{placeins}
\usepackage{stfloats}
\usepackage{tikz}
\usepackage{xcolor}
\usepackage{url}
\usetikzlibrary{arrows.meta,positioning}

\graphicspath{{figures/}}

\usepackage[hidelinks]{hyperref}
\hypersetup{
    colorlinks=false,
    linkcolor=false,
    filecolor=magenta,      
    urlcolor=cyan,
    citecolor=green
}

\begin{document}

\title{Rapid Earthquake-to-Tsunami Waveform Generation via Large-Scale Multi-GPU FFT Convolution Applied to the Cascadia Subduction Zone}

\author{
\IEEEauthorblockN{Bowen Shi}
\IEEEauthorblockA{\textit{Oden Institute for Computational Engineering and Sciences} \\
\textit{The University of Texas at Austin}\\
Austin, TX, USA \\
bowenshi@utexas.edu}
\and
\IEEEauthorblockN{Sreeram Venkat}
\IEEEauthorblockA{\textit{Oden Institute for Computational Engineering and Sciences} \\
\textit{The University of Texas at Austin}\\
Austin, TX, USA \\
srvenkat@utexas.edu}
\and
\IEEEauthorblockN{Stefan Henneking}
\IEEEauthorblockA{\textit{Oden Institute for Computational Engineering and Sciences} \\
\textit{The University of Texas at Austin}\\
Austin, TX, USA \\
stefan@oden.utexas.edu}
\and
\IEEEauthorblockN{Omar Ghattas}
\IEEEauthorblockA{\textit{Oden Institute for Computational Engineering and Sciences,} \\
\textit{Walker Department of Mechanical Engineering} \\
\textit{The University of Texas at Austin}\\
Austin, TX, USA \\
omar@oden.utexas.edu}
}

\maketitle

\begin{abstract}
Data-driven methods for earthquake and tsunami early warning rely on large ensembles of rupture scenarios and their resulting waveforms, but generating such datasets with repeated high-fidelity seismic and tsunami simulations is prohibitively expensive. We exploit the linear time-invariant structure of both dynamics to precompute elastic Green's functions and acoustic-gravity adjoint responses, reducing the source-to-waveform map to two consecutive convolution operators. We evaluate these convolutions with a distributed, FFT-accelerated GPU pipeline that partitions the large seafloor grid across GPUs and directly generates the final observation waveforms. We demonstrate the scalability of this pipeline for the Cascadia Subduction Zone with 963 subfaults, 2,416,530 seafloor grid points, 64 observation locations, and 256 timesteps, requiring 9.45~TiB of aggregate GPU memory. On 64 GB200 GPUs within one NVL72 domain, the pipeline generates waveforms in 24 ms per rupture once the response operators are resident, enabling large rupture ensembles to be evaluated within minutes.
\end{abstract}

\begin{IEEEkeywords}
Data-driven prior modeling,
Green's functions,
Linear time-invariant dynamical systems,
Earthquake rupture,
Cascadia Subduction Zone,
Block Toeplitz matrix
\end{IEEEkeywords}

\section{Introduction}

Megathrust earthquakes and the resulting tsunamis pose significant risks to coastal communities, infrastructure, and human life. Near a subduction zone, destructive waves may reach adjacent
coastlines within minutes, severely limiting the time to generate and issue forecasts.
Conventional earthquake-based tsunami warning systems first use seismic observations to rapidly estimate source parameters, particularly moment magnitude and hypocentral location~\cite{Hirshorn2021, Kamigaichi2009}. These estimates provide an initial description of the event but do not
resolve the heterogeneous slip and rupture evolution that determine the
time-dependent motion of the seafloor. Offshore seafloor pressure sensors
provide direct measurements of the evolving ocean response and can therefore
be used to better constrain the tsunami source~\cite{leveque2018cascadia, venkat2026oed}. Recent work
used these observations to develop a high-fidelity digital twin for tsunami early warning applied to the
Cascadia Subduction Zone (CSZ), formulating the inference of seafloor motion
and the subsequent forecasting of tsunami wave heights as a large-scale
Bayesian inverse problem \cite{henneking2026real, henneking2026goal, henneking2025bell}.

In the Bayesian inverse formulation of~\cite{henneking2026goal}, the unknown spatiotemporal seafloor motion is assigned a zero-mean Gaussian prior. Its covariance matrix is block-diagonal in time, with each temporal block given by the same Matérn-type spatial covariance that is generated from an inverse elliptic differential operator \cite{lindgren2011explicit}. This spatial covariance suppresses rough components of the seafloor motion, reflecting the regularizing role of Gaussian priors in Bayesian inverse problems~\cite{ghattas2021learning, stuart2010inverse}. Treating the temporal blocks as independent and identical also preserves temporal shift invariance and, consequently, the block Toeplitz structure required by the fast inference algorithms~\cite{henneking2026goal}. Although computationally advantageous, this prior does not encode the spatiotemporal correlations produced by heterogeneous fault slip, rupture onset and propagation, and seismic wave dynamics. To incorporate this earthquake-driven structure into prior models and data-driven tsunami forecasting, large ensembles of physically plausible earthquake realizations and their resulting waveforms are needed.

\begin{figure}[htb]
    \centering
    \includegraphics[width=\columnwidth]
    {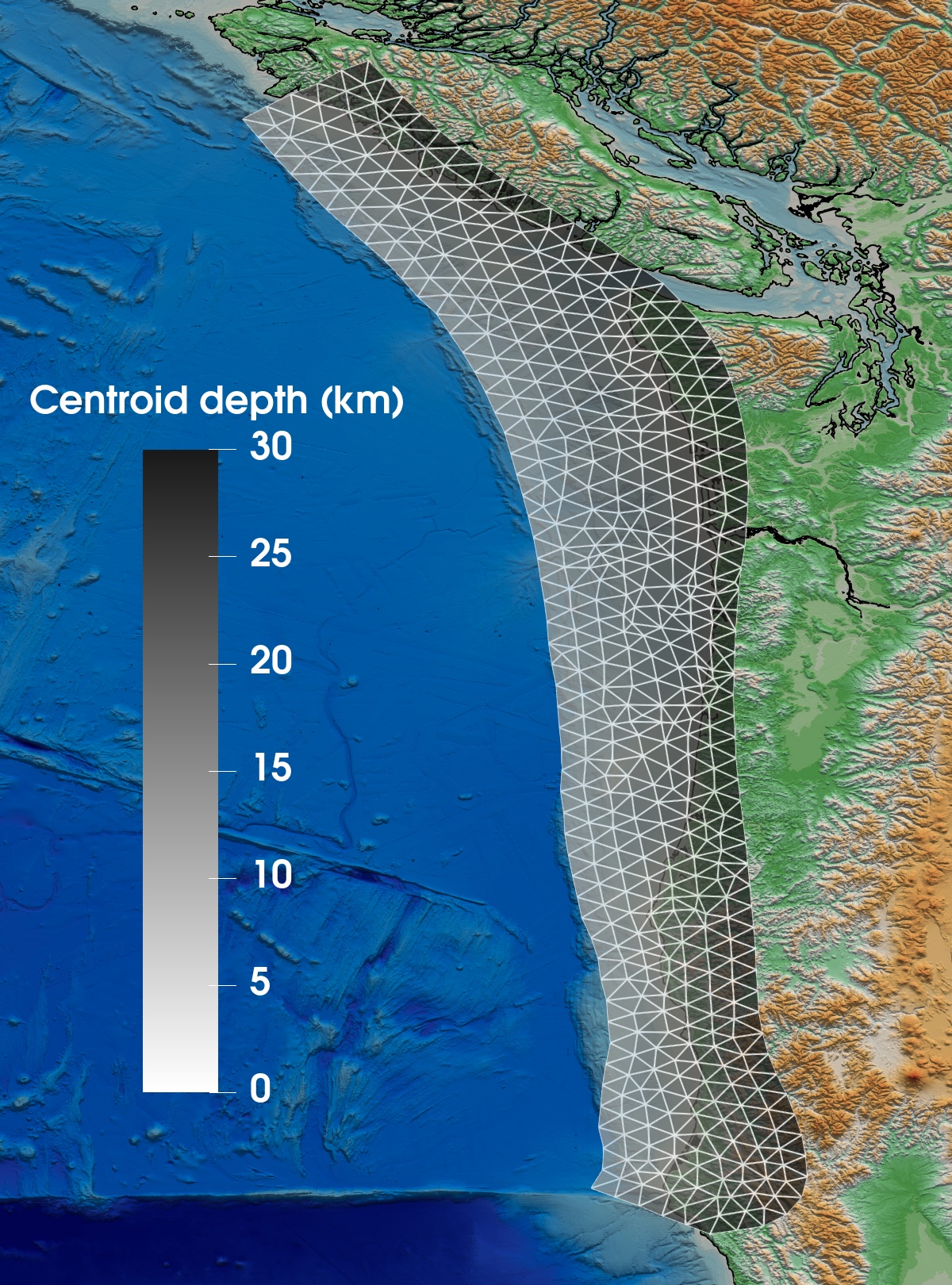}
    \caption{CSZ fault-plane discretization used for the FakeQuakes rupture realizations, overlaid with the topobathymetry of the Cascadia region. The 963 triangular subfaults are shaded by centroid depth, the vertical depth of the subfault's geometric center, in kilometers below sea level.}
    \label{fig:fault-patches}
\end{figure}

Because great tsunamigenic earthquakes are rare, these ensembles must rely
largely on synthetic ruptures. Previous data-driven tsunami forecasting
studies have used large sets of earthquake scenarios generated from
stochastic source models (e.g., \cite{makinoshima2021, mulia2022, leveque2016generating, williamson2020}). For the CSZ,
FakeQuakes \cite{melgar2016} generates such scenarios on a fault-plane discretization of
triangular subfaults, shown in
Figure~\ref{fig:fault-patches}. For a prescribed magnitude, FakeQuakes uses
empirical scaling relations to sample a spatially correlated slip field and
assigns a hypocenter, rupture-onset time, and rise time. These quantities
determine a slip-rate function on each subfault. A FakeQuakes realization
therefore describes the evolution of slip on the fault, but not the seafloor motion or 
offshore pressure waveforms generated by the event.

{Two sequential physical processes connect these fault slip-rate functions
to offshore pressure and tsunami waveform observations. First, slip on the fault radiates elastic
waves through the solid Earth and produces time-dependent seafloor
displacement. Second, the seafloor motion excites acoustic--gravity waves in
the ocean, producing tsunami waves and pressure signals at offshore sensors. The seafloor
motion is the output of the solid-Earth model and the input to the ocean
model. High-fidelity numerical models describe these two processes with
elastodynamic and acoustic--gravity wave equations. Solving these models
separately for tens of thousands of earthquake scenarios, even with highly optimized 
implementations (e.g., \cite{henneking2025bell, uphoff2017sumatra, abrahams2023comparison, tu2026tensor}), 
would make ensemble generation prohibitively expensive.}
This leads to the central question of
this work: \textit{How can we generate offshore observation waveforms for
large ensembles of plausible Cascadia ruptures without computing wave
propagation via expensive-to-solve partial differential equations (PDEs) 
separately for every realization?}

Both models in this workflow are linear time-invariant (LTI), so their impulse
responses can be precomputed and reused across earthquake realizations.
Elastic Green's functions map the fault slip-rate functions to seafloor
displacement, while coupled ocean acoustic and tsunami wave dynamics, 
computed via adjoint acoustic--gravity PDE solutions 
\cite{lotto2015tsunami, henneking2025bell}, map the seafloor
motion to pressure observations and tsunami wave heights. The complete
source-to-waveform map therefore consists of two consecutive convolutions,
which form block lower-triangular Toeplitz matrix--vector multiplications (matvecs) after time discretization.
The FFTMatvec library \cite{venkat2025fft} uses temporal FFTs to transform these convolutions into independent
frequency-domain matvecs, while the large seafloor dimension
is partitioned across GPUs. This makes the computation
highly parallel but requires the frequency-domain response kernels to remain
in GPU memory; at full CSZ scale, these kernels occupy several terabytes. Our
distributed implementation keeps the intermediate seafloor field on the GPUs
between the two convolutions and stores only the final observation waveforms.
Our contributions are as follows:
\begin{itemize}
  \item \textbf{Multi-GPU, FFT-accelerated source-to-waveform generation.}
        By exploiting the LTI structure of the earthquake
        and tsunami dynamics, we precompute elastic Green's functions and
        acoustic--gravity adjoint responses. We formulate the resulting
        source-to-waveform map as two consecutive block lower-triangular
        Toeplitz matvecs and evaluate them using distributed FFT-based
        algorithms on GPUs. The transient seafloor field remains resident
        on the GPUs and is passed directly between the two operators,
        avoiding repeated wave-equation solves and external storage of the
        full space--time field.

  \item \textbf{Application to tsunami waveform generation on the CSZ.}
        We demonstrate the source-to-waveform pipeline on a full-scale CSZ
        configuration with 963 subfaults, 2,416,530 seafloor grid points, 64
        observation locations, and 256 timesteps. The
        frequency-domain Green's functions and adjoint vectors require
        approximately 9.45~TiB of aggregate resident GPU memory. Using 64
        GB200 GPUs within an NVL72 domain, the pipeline evaluates one rupture
        in 24~ms and delivers a $21.8\times$ speedup over a distributed
        PyTorch convolution baseline implementation.
\end{itemize}

\section{Methods}

\subsection{Stochastic ruptures and slip-rate functions}

The Cascadia fault interface is discretized into 963 triangular subfaults
(Fig.~\ref{fig:fault-patches}). FakeQuakes first selects a rupture extent and
hypocenter for a prescribed target magnitude. It constructs an anisotropic
von K\'arm\'an slip covariance, samples it with a truncated Karhunen--Lo\`eve
expansion, and assigns slip to the active subfaults \cite{melgar2016}. Our
target is to generate tsunami waveforms for tens of thousands of these
stochastic ruptures.

For each subfault, the sampled slip, rupture-onset time, 
and rise time uniquely determine its slip-rate function. We
use the dip-slip component and only compute the resulting vertical seafloor
displacement which causes the tsunami. The temporal domain is discretized with 
\(N_t=256\) one-second timesteps.

\subsection{Impulse-response operators for LTI dynamics}

The kinematic rupture model for seismic wave propagation and the acoustic--gravity model for tsunami wave propagation are both governed by LTI dynamics. A generic LTI system for unknown $\boldsymbol{u}$ may be written as
\begin{equation*}
    \frac{\partial \boldsymbol{u}}{\partial t}
    =
    \mathcal{A}\boldsymbol{u}
    +\mathcal{C}\boldsymbol{x},
    \qquad
    \boldsymbol{y}
    =
    \mathcal{B}\boldsymbol{u},
    \label{eq:lti-system}
\end{equation*}
with appropriate initial and boundary conditions. The operators
\(\mathcal{A}\), \(\mathcal{B}\), and \(\mathcal{C}\) are linear and
independent of time, while \(\boldsymbol{x}\) and \(\boldsymbol{y}\) denote
the input and output, respectively. After time discretization, the dynamics
take the form
\begin{equation*}
    \mathbf{u}_{n}
    =
    \mathbf{A}\mathbf{u}_{n-1}
    +\mathbf{C}\mathbf{x}_n,
    \qquad
    \mathbf{y}_{n}
    =
    \mathbf{B}\mathbf{u}_{n}.
\end{equation*}
For homogeneous initial conditions, eliminating the state gives
\begin{equation*}
    \mathbf{y}_{n}
    =
    \sum_{k=0}^{n}
    \mathbf{B}\mathbf{A}^{k}\mathbf{C}\,
    \mathbf{x}_{n-k}.
    \label{eq:lti-impulse-sum}
\end{equation*}
The input--output map is therefore determined by the impulse-response blocks
\(\mathbf{B}\mathbf{A}^{k}\mathbf{C}\). Time invariance makes these blocks
depend only on the time lag, while causality excludes responses at negative
lags. We construct such impulse responses offline for the elastic and
acoustic--gravity equations and reuse them for every rupture realization. 

As the number of seafloor grid points $N_m$ is significantly larger than the number of subfaults $N_f$ and the number of observation points $N_q$, the two offline constructions therefore proceed in complementary directions:
each elastic forward calculation fixes a fault source and produces its
response at all seafloor points, whereas each acoustic--gravity adjoint
calculation fixes an output channel and produces its response to motion at
all seafloor input points. Thus, a total of $N_f$ elastodynamical Green's functions and $N_q$ adjoint acoustic--gravity responces are derived.

\subsubsection{Elastic Green's functions}

The first operator maps slip-rate functions on the fault to vertical seafloor
displacement. We construct its impulse responses with the
frequency--wavenumber (FK) method \cite{zhu2002note} using its Python implementation \cite{xi2021pyfk}. 
A layered velocity field is assumed in the seismic wave propagation. 
For a point source
\(\boldsymbol{\xi}\), Fourier transformation in time and in the horizontal
coordinates reduces the layered-elastodynamic equations to independent depth
problems for each angular frequency \(\omega\) and horizontal wavenumber
\(\kappa\). For each pair \(\omega, \kappa\), the FK method
solves an independent set of layered-medium ordinary differential equations (ODEs), which produces spectral coefficients
\(K_{\nu}(\kappa,\omega)\), with $\nu=0, 1, 2$. For a given seafloor grid point $i$, let
\begin{equation*}
    \widehat g_z(r_i,\phi_i,\omega)
    =
    \sum_{\nu=0}^{2} b_\nu(\phi_i)
    \int_{0}^{\infty}
    K_\nu(\kappa,\omega)J_\nu(\kappa r_i)\,\kappa\,\mathrm{d}\kappa ,
    \label{eq:fk-hankel}
\end{equation*}
where \(r_i\) and \(\phi_i\) are the source--receiver distance and azimuth,
respectively, \(b_\nu(\phi_i)\) is the azimuth-dependent coefficient and \(J_\nu\) is the Bessel function of order \(\nu\). The corresponding time-domain
point-source Green's function for vertical displacement $g_z(\mathbf x_i,\boldsymbol{\xi},t)$ can be obtained by an inverse Fourier transform of $\widehat g_z(\mathbf x_i,\boldsymbol{\xi},\omega)$.

The ODE solves are independent across \((\kappa,\omega)\) pairs and are therefore
embarrassingly parallel; for a fixed source, the coefficients
\(K_\nu(\kappa,\omega)\) are shared by all receivers; only the \(b_\nu(\phi_i)\) and \(J_\nu(\kappa r_i)\) vary between receivers. We therefore
compute the coefficients once, retain them on the GPU, and reconstruct the
vertical dip-slip response for many seafloor receivers in parallel batches.

The fault unknown is uniform slip over a triangular subfault rather than a
point source. For a triangular subfault $T_j$, the Green's function with 
respect to seafloor grid point $i$ reads:
\begin{equation*}
\begin{aligned}
  {g}_{i,j}[t]
    &=
    \frac{1}{|T_j|}
    \int_{T_j}
    g_z(\boldsymbol{x}_i,\boldsymbol{\xi},t)\,
    \mathrm{d}A(\boldsymbol{\xi}) 
\end{aligned}
\label{eq:triangle-gf-average}
\end{equation*}
In computations, we approximate this area-averaged impulse
response using a 16-point quadrature rule. 
Thus, one forward Green's function construction per quadrature source
produces its response at all seafloor receivers. We retain only the dip-slip component and compute the resulting vertical seafloor displacement, which drives the tsunami. Define
\begin{equation*}
    \left[\mathbf{F}_{G,k}\right]_{i,j}
    =
  {g}_{i,j}[t_k],
    \qquad
    \mathbf{F}_{G,k}
    \in
    \mathbb{R}^{N_m\times N_f}.
    \label{eq:elastic-block-entry}
\end{equation*}
 Writing
\(\mathbf{m}_n\in\mathbb{R}^{N_m}\) for the displacement and \(\mathbf{s}_n\in\mathbb{R}^{N_f}\) for the slip-rate function at timestep \(n\), we have
\begin{equation}
    \mathbf{m}_n
    =
    \sum_{k=0}^{n}
    \mathbf{F}_{G,k}\mathbf{s}_{n-k}.
    \label{eq:elastic-response}
\end{equation}

\subsubsection{Acoustic--gravity adjoint responses}

The second operator maps the seafloor velocity field to seafloor-pressure
or tsunami-height waveforms. 
From seafloor displacement \(\mathbf m\), we can compute velocity 
\begin{equation*}
    \mathbf v = \mathbf D_t\mathbf m,
\end{equation*}
where \(\mathbf D_t\) is the discrete temporal-difference stencil applied
independently at every seafloor point. At time step \(n\), the waveform
response is
\begin{equation}
    \mathbf q_n
    =
    \sum_{k=0}^{n}
    \mathbf F_{q,k}\mathbf v_{n-k},
    \qquad
    \mathbf F_{q,k}\in\mathbb R^{N_q\times N_m}.
    \label{eq:waveform-response}
\end{equation}

Constructing \(\mathbf F_{q,k}\) with forward simulations would require one
acoustic--gravity PDE solve for each of the \(N_m\) seafloor input points.
Each PDE solution produces one column of the response blocks, which could be expensive if $N_m$ is large. We instead use
the adjoint formulation of the PDEs
\cite{henneking2026goal}. One adjoint solve using a point
source associated with an observation location recovers the corresponding row of
\(\mathbf F_{q,k}\). The complete operator consequently requires \(N_q\)
adjoint solves rather than \(N_m\) forward solves. For the CSZ discretization with \(N_m=2{,}416{,}530\), a single subfault layered-medium Green's-function forward solve requires approximately one hour on 4 NVIDIA A100 GPUs, while a single adjoint acoustic--gravity wave propagation requires approximately one hour on 512 NVIDIA A100 GPUs~\cite{henneking2025bell}.

\subsubsection{Block Toeplitz representation}

For \(\alpha\in\{G,q\}\), stacking all timesteps in
Equations~\eqref{eq:elastic-response} and \eqref{eq:waveform-response} gives the
block lower-triangular Toeplitz operator
\begin{equation*}
\begin{aligned}
    \mathbf{F}_\alpha
    &=
    \operatorname{BLTT}\!\left(
        \mathbf{F}_{\alpha,0},
        \ldots,
        \mathbf{F}_{\alpha,N_t-1}
    \right), \\
    &=
    \begin{bmatrix}
        \mathbf{F}_{\alpha,0}
        & 0
        & \cdots
        & 0 \\
        \mathbf{F}_{\alpha,1}
        & \mathbf{F}_{\alpha,0}
        & \ddots
        & \vdots \\
        \vdots
        & \ddots
        & \ddots
        & 0 \\
        \mathbf{F}_{\alpha,N_t-1}
        & \cdots
        & \mathbf{F}_{\alpha,1}
        & \mathbf{F}_{\alpha,0}
    \end{bmatrix}.
\end{aligned}
\label{eq:block-toeplitz}
\end{equation*}
Here, \(\operatorname{BLTT}(\cdot)\) denotes the block
lower-triangular Toeplitz matrix generated by a sequence of impulse-response
blocks \cite{venkat2025fft, venkat2025mixed}.
Then,
\[
    \mathbf{F}_G:
    \mathbb{R}^{N_fN_t}
    \longrightarrow
    \mathbb{R}^{N_mN_t}
\]
maps the subfault slip-rate functions to seafloor motions, and 
\[
    \mathbf{F}_q:
    \mathbb{R}^{N_mN_t}
    \longrightarrow
    \mathbb{R}^{N_qN_t}
\]
maps seafloor motion to the output
waveforms.
The large seafloor motion dimension \(N_mN_t\) is the output dimension of
\(\mathbf{F}_G\) and the input dimension of \(\mathbf{F}_q\). This shared
dimension permits the opposite row and column partitions introduced in the
next subsection, allowing the intermediate seafloor field to remain local
and GPU resident between the two distributed Toeplitz matvecs.

\subsection{Distributed FFT convolution}

The two response operators share the seafloor dimension, along which the
distributed decomposition is defined. Let
\(\{\mathcal I_r\}_{r=0}^{P-1}\) be a disjoint partition of the \(N_m\)
seafloor points, and let \(N_m^{(r)}=|\mathcal I_r|\). At every temporal lag
\(k\), rank \(r\) stores
\begin{equation*}
\mathbf F_{G,k}^{(r)}
=
\mathbf F_{G,k}[\mathcal I_r,:],
\qquad
\mathbf F_{q,k}^{(r)}
=
\mathbf F_{q,k}[:,\mathcal I_r],
\label{eq:opposite-partitions}
\end{equation*}
where
\[
\mathbf F_{G,k}^{(r)}
\in\mathbb R^{N_m^{(r)}\times N_f},
\qquad
\mathbf F_{q,k}^{(r)}
\in\mathbb R^{N_q\times N_m^{(r)}}.
\]
Thus, the elastic response blocks are partitioned by their seafloor output
rows, while the waveform response blocks are partitioned by the matching
seafloor input columns.

These lag-dependent blocks generate the local block Toeplitz operators
\begin{equation*}
\begin{aligned}
\mathbf F_G^{(r)}
&=
\operatorname{BLTT}\!\left(
\mathbf F_{G,0}^{(r)},\ldots,
\mathbf F_{G,N_t-1}^{(r)}
\right),\\
\mathbf F_q^{(r)}
&=
\operatorname{BLTT}\!\left(
\mathbf F_{q,0}^{(r)},\ldots,
\mathbf F_{q,N_t-1}^{(r)}
\right).
\end{aligned}
\label{eq:local-toeplitz-operators}
\end{equation*}
They have dimensions
\[
\mathbf F_G^{(r)}
\in
\mathbb R^{N_m^{(r)}N_t\times N_fN_t},
\qquad
\mathbf F_q^{(r)}
\in
\mathbb R^{N_qN_t\times N_m^{(r)}N_t}.
\]

For a slip-rate function \(\mathbf s\), rank \(r\) first computes its
local seafloor motion
\(\mathbf m^{(r)}=\mathbf F_G^{(r)}\mathbf s\).
Because \(\mathbf D_t\) acts independently on each seafloor grid point, the complete
distributed composition can be written compactly as
\begin{equation}
\mathbf q
=
\sum_{r=0}^{P-1}
\mathbf F_q^{(r)}
\mathbf D_t
\mathbf F_G^{(r)}
\mathbf s.
\label{eq:distributed-composition}
\end{equation}
The output partition of \(\mathbf F_G\) therefore matches the input partition
of \(\mathbf F_q\). The seafloor field remains distributed and
on-device between the two operator applications. Only the compact slip-rate functions are
broadcast, and only the output waveforms are summed across ranks.
\begin{algorithm}[htb]
\caption{Distributed source-to-waveform evaluation.}
\label{alg:distributed-pipeline}
\begin{algorithmic}[1]
\Require Slip-rate function \(\mathbf s\); local operators
\(\mathbf F_G^{(r)}\) and \(\mathbf F_q^{(r)}\).
\Ensure Waveforms \(\mathbf q\)
\State \(\mathbf s\gets\Call{Broadcast}{\mathbf s}\)
\State \(\mathbf m^{(r)}
       \gets
       \mathbf F_G^{(r)}\mathbf s\)
\Comment{Local matvec in freq or time}
\State \(\mathbf v^{(r)}\gets\mathbf D_t\mathbf m^{(r)}\)
\Comment{Local displacement to velocity}
\State \(\mathbf q^{(r)}
       \gets\mathbf F_q^{(r)} \mathbf v^{(r)}\)
\Comment{Local matvec in freq or time}
\State \(\mathbf q
       \gets\Call{ReduceSum}{\mathbf q^{(r)}}\)
\State \Return \(\mathbf q\)
\end{algorithmic}
\end{algorithm}

Algorithm~\ref{alg:distributed-pipeline} implements the distributed strategy from \eqref{eq:distributed-composition}. For either
\(\mathbf F_G^{(r)}\) or \(\mathbf F_q^{(r)}\), the local matvec represents
the causal convolution. We apply both local block Toeplitz operators using
FFTMatvec \cite{venkat2025fft} in frequency space. To quantify the benefit of the frequency-domain application, we
also implement a direct time-domain convolution baseline. The two
implementations use the same response operators, data decomposition, and
communication pattern, and differ only in how each local Toeplitz matvec 
within the block Toeplitz structure is evaluated.
 FFTMatvec applies each local Toeplitz matvec through Fourier diagonalization, 
 whereas the time-domain
reference evaluates the equivalent temporal convolutions directly. Their
performance comparison isolates the local matvec implementation while
holding the operator, data distribution, and communication pattern fixed.

\subsubsection{Frequency-domain convolution with FFTMatvec}
For a BLTT matvec \(\mathbf y=\mathbf F\mathbf x\), FFTMatvec embeds $\mathbf F$
into a block-circulant matrix and diagonalizes
the temporal dimension using the Fourier transform. After zero-padding to
length \(2N_t\), the convolution becomes
\begin{equation*}
\widehat{\mathbf y}_{\ell}
=
\widehat{\mathbf F}_{\ell}
\widehat{\mathbf x}_{\ell},
\qquad
\ell=0,\ldots,N_t,
\end{equation*}
where \(\widehat{\mathbf F}_{\ell}\) is the complex response block at
frequency bin \(\ell\) after temporal Fourier transform.

The Fourier transformed response blocks
\(\{\widehat{\mathbf F}_{\ell}\}_{\ell=0}^{N_t}\) are computed once during
setup and remain resident in GPU memory. Algorithm~\ref{alg:fftmatvec-local} summarizes
the FFTMatvec operation.

\begin{algorithm}[htb]
\caption{Frequency-domain BLTT FFT-based matvec.}
\label{alg:fftmatvec-local}
\begin{algorithmic}[1]
\Require Input \(\mathbf x\); frequency-domain response blocks
\(\{\widehat{\mathbf F}_{\ell}\}_{\ell=0}^{N_t}\)
\Ensure Output \(\mathbf y\)
\State \(\widetilde{\mathbf x}
       \gets \Call{ZeroPad}{\mathbf x,2N_t}\)
\State \(\widehat{\mathbf x}
       \gets \Call{FFT}{\widetilde{\mathbf x}}\)
\State \(\widehat{\mathbf y}_{\ell}
       \gets \widehat{\mathbf F}_{\ell}
       \widehat{\mathbf x}_{\ell}\),
       \(\ell=0,\ldots,N_t\), in parallel
\State \(\widetilde{\mathbf y}
       \gets \Call{IFFT}{\widehat{\mathbf y}}\)
\State \(\mathbf y
       \gets \Call{Truncate}{\widetilde{\mathbf y},N_t}\)
\State \Return \(\mathbf y\)
\end{algorithmic}
\end{algorithm}
\subsubsection{Direct time-domain reference}

The reference implementation computes the same local product
\(\mathbf y=\mathbf F\mathbf x\), but evaluates it directly in the time
domain. Let
\[
f_{ij}[k]=[\mathbf F_k]_{i,j}
\]
denote the temporal response from input channel \(j=1,\ldots,N_{\mathrm {in}}\) to output channel
\(i\). For $n=0,\ldots,N_t-1,$ we have
\begin{equation}
y_i[n]
=
\sum_{j=1}^{N_{\mathrm{in}}}
\left(f_{ij}\ast_t x_j\right)[n]
=
\sum_{j=1}^{N_{\mathrm{in}}}
\sum_{k=0}^{n}
f_{ij}[k]\,x_j[n-k]\label{eq:time-domain-conv}
\end{equation}
where \(\ast_t\) denotes one-dimensional temporal convolution. Thus, the
direct implementation evaluates one temporal convolution for each
input--output channel pair and sums the results over the input channels.

In our reference implementation, we compute \eqref{eq:time-domain-conv} using PyTorch's
\texttt{torch.nn.functional.conv\_transpose1d} API
\cite{NEURIPS2019_bdbca288}, which evaluates the pairwise one-dimensional convolutions and
sums them over the input-channel dimension.

\section{Results}

We evaluate the source-to-waveform pipeline at the full CSZ scale, using 300 m grid resolution and study its distributed performance as the GPU count increases.
The experiments compare FFTMatvec with a direct time-domain PyTorch reference
and then apply both implementations to the complete CSZ operator.

\subsection{Systems and scaling configuration}

The experiments use two distinct GPU partitions of TACC's Vista system. 
Each node of the GH200 partition has one NVIDIA Grace--Hopper Superchip 
with 96~GiB HBM3 memory, connected via 400~Gb/s NDR InfiniBand. 
The GB200 partition has 18 compute nodes, each comprised of two NVIDIA 
Grace--Blackwell Superchips that each connect two Blackwell GPUs and a Grace CPU.
The GB200 nodes are connected through a fifth-generation NVLink Switch fabric as a 
single 72-GPU NVL72 domain. Each GPU provides 184~GiB of measured usable 
HBM3E and 1.8~TB/s of bidirectional GPU-to-GPU NVLink bandwidth. 
We use one MPI rank per GPU, i.e., one rank per GH200 and four ranks per GB200 node.

Our PyTorch baseline directly calls
\texttt{torch.nn.functional.conv\_transpose1d} for both time-domain
  convolutions. A convolution is issued as a single call whenever its local
  tensor shape is supported; otherwise, only the seafloor grid dimension is divided
  using the largest power-of-two tile satisfying its cuDNN backend's limit.
  The tile is determined once and fixed for all reported runs. For FFTMatvec 
  and the PyTorch baseline, we fix the software stack as 
PyTorch 2.9.1, CUDA 12.8, cuDNN 9.10.2, and NCCL 2.27.5. 
All computations are done in FP64.

\begin{table}[htb]
\caption{Scaling configurations on Vista GH200 and GB200 NVL72\\
for \(P\in\{8,16,32,64\}\) GPUs.}
\label{tab:experiments}
\centering
\footnotesize
\setlength{\tabcolsep}{4pt}
\renewcommand{\arraystretch}{1.1}
\begin{tabular}{@{}lccc@{}}
\toprule
Study & Subfaults & Observation locations & Seafloor grid points \\
\midrule
Strong scaling & 60 & 4 &
\(2{,}416{,}530\) \\
Weak scaling & 480&32 &
\(\left\lfloor 2{,}416{,}530P/64\right\rfloor\) \\
\bottomrule
\end{tabular}
\end{table}

The distributed dimension in both operators is the seafloor grid.
Our strong scaling experiments measure how additional GPUs reduce the time to apply a fixed
source-to-waveform map. We retain all \(N_m=2{,}416{,}530\) seafloor points
and \(N_t=256\) timesteps, but select 60 of the 963 subfault Green's functions and four
of the 64 waveform observation locations. This is the largest fixed problem that
fits on eight GH200 GPUs. The same data are partitioned over 8, 16, 32, and
64 GPUs on both machines.

In the weak scaling experiments, the local spatial work is held constant at 
approximately 37,758 seafloor points per GPU; the global grid size is increased 
from 302,066 points on 8 GPUs to the full 2,416,530 points on 64 GPUs. 
The fixed operator contains 480 subfault Green's functions and 32 waveform locations.
The configurations used for strong and weak scaling are summarized in Table~\ref{tab:experiments}.

After operator setup and warm-up, we report the average end-to-end time over
64 pipelined ruptures throughout the experiments. The timing includes per-rupture I/O, broadcast,
convolution, finite difference, and reduction
(Algorithm~\ref{alg:distributed-pipeline}). The I/O for small source-time function input and waveform output is overlapped with computation in the pipeline and has negligible overhead.

\subsection{Multi-GPU scaling results}

\begin{figure*}[htb]
\centering
\includegraphics[width=0.8\textwidth]{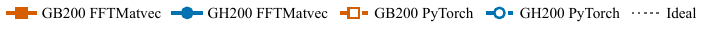}\\
\vspace{2pt}
\includegraphics[width=0.49\textwidth]{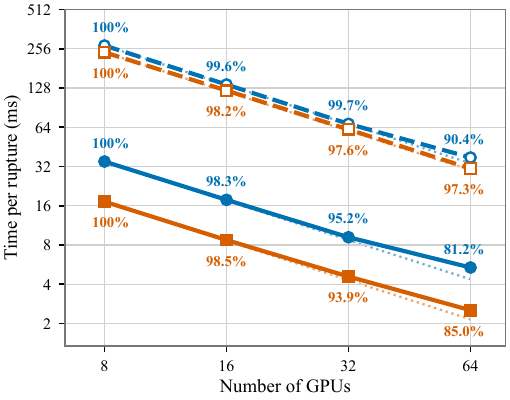}
\hfill
\includegraphics[width=0.49\textwidth]{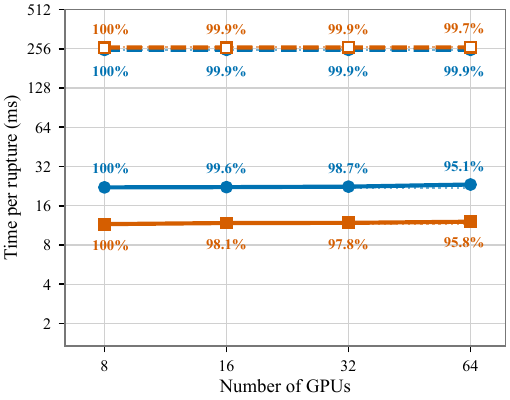}
\caption{Multi-GPU strong scalability (left) and weak scalability (right) results. 
The strong-scaling global problem size is fixed at 60 subfaults, 4 observation locations, and \(N_m=2{,}416{,}530\) seafloor grid points. 
The weak-scaling per-GPU problem size is fixed at 480 subfaults, 32 observation locations, and 37,758 seafloor grid points. 
The text labels denote the parallel efficiencies relative to the corresponding eight-GPU baseline performance. 
}
\label{fig:scaling}
\end{figure*}

Figure~\ref{fig:scaling} shows the strong and weak scalability results, 
including parallel efficiencies, for the multi-GPU source-to-waveform pipeline. 

For the fixed global problem in strong scaling, FFTMatvec achieves speedups
of \(6.5\times\) on GH200 and \(6.8\times\) on GB200 when increasing the
allocation from 8 to 64 GPUs. At 64 GPUs, this corresponds to parallel
efficiencies of \(81.2\%\) and \(85.0\%\), respectively. Across all GPU counts, FFTMatvec on GB200 is
approximately \(2\times\) faster than on GH200. FFTMatvec reduces the time-to-solution 
by approximately \(7\times\) relative to PyTorch on GH200 and by
\(12\)--\(14\times\) on GB200. Although our PyTorch baseline exhibits higher
parallel efficiency, its runtime remains substantially larger because the
local direct-convolution kernels dominate the computation.

The weak scaling experiment shows that FFTMatvec also
scales effectively as the global problem size increases. From 8 to 64 GPUs,
FFTMatvec retains weak-scaling efficiencies of \(95.1\%\) on GH200 and
\(95.8\%\) on GB200, while GB200 remains approximately \(1.9\times\) faster.

We note that in the PyTorch weak scaling case, results on GB200 are approximately \(4\%\)
slower than on GH200 despite using the same software stack and tensor shapes. CUDA-event timers for each stage in Algorithm~\ref{alg:distributed-pipeline} show that the first convolution takes approximately \(252\)~ms on GB200 and \(241\)~ms on GH200, while the second convolution and final reduce sum are slightly faster on GB200. 
We attribute these performance numbers to the architecture-specific efficiency of cuDNN's FP64 convolution kernels for this tensor geometry.

\subsection{Application to CSZ}
Finally, we use the complete operator with 963 subfaults, 64 observation locations,
and the full \(2{,}416{,}530\)-point CSZ seafloor grid and perform experiments on 64 GB200 GPUs on one NVL72 domain. For each FakeQuakes rupture, the pipeline returns 64 waveforms with 256 timesteps.

\begin{figure*}[hp]
    \centering
    \includegraphics[width=0.33\textwidth]
    {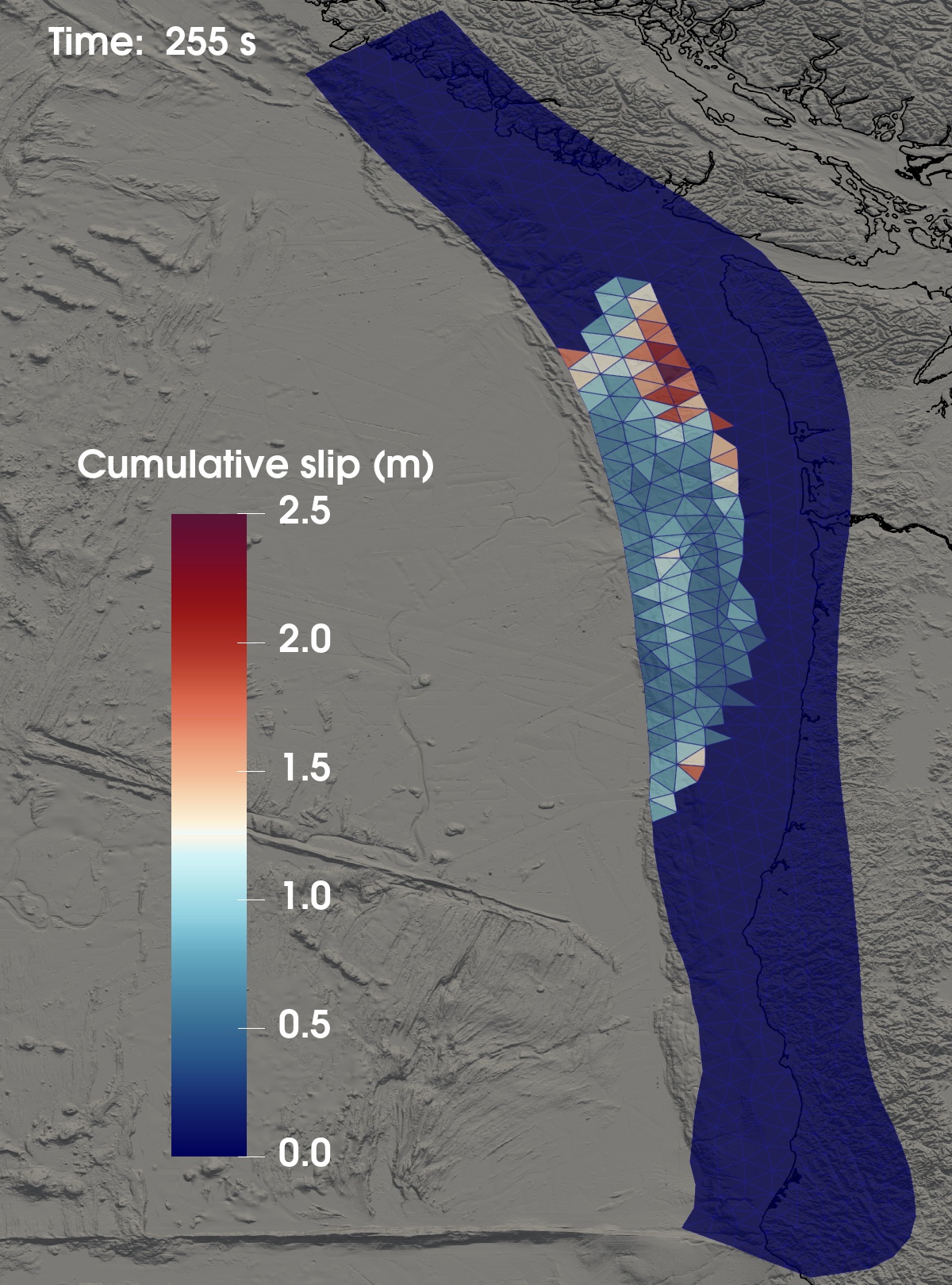}%
    \hfill%
    \includegraphics[width=0.33\textwidth]
    {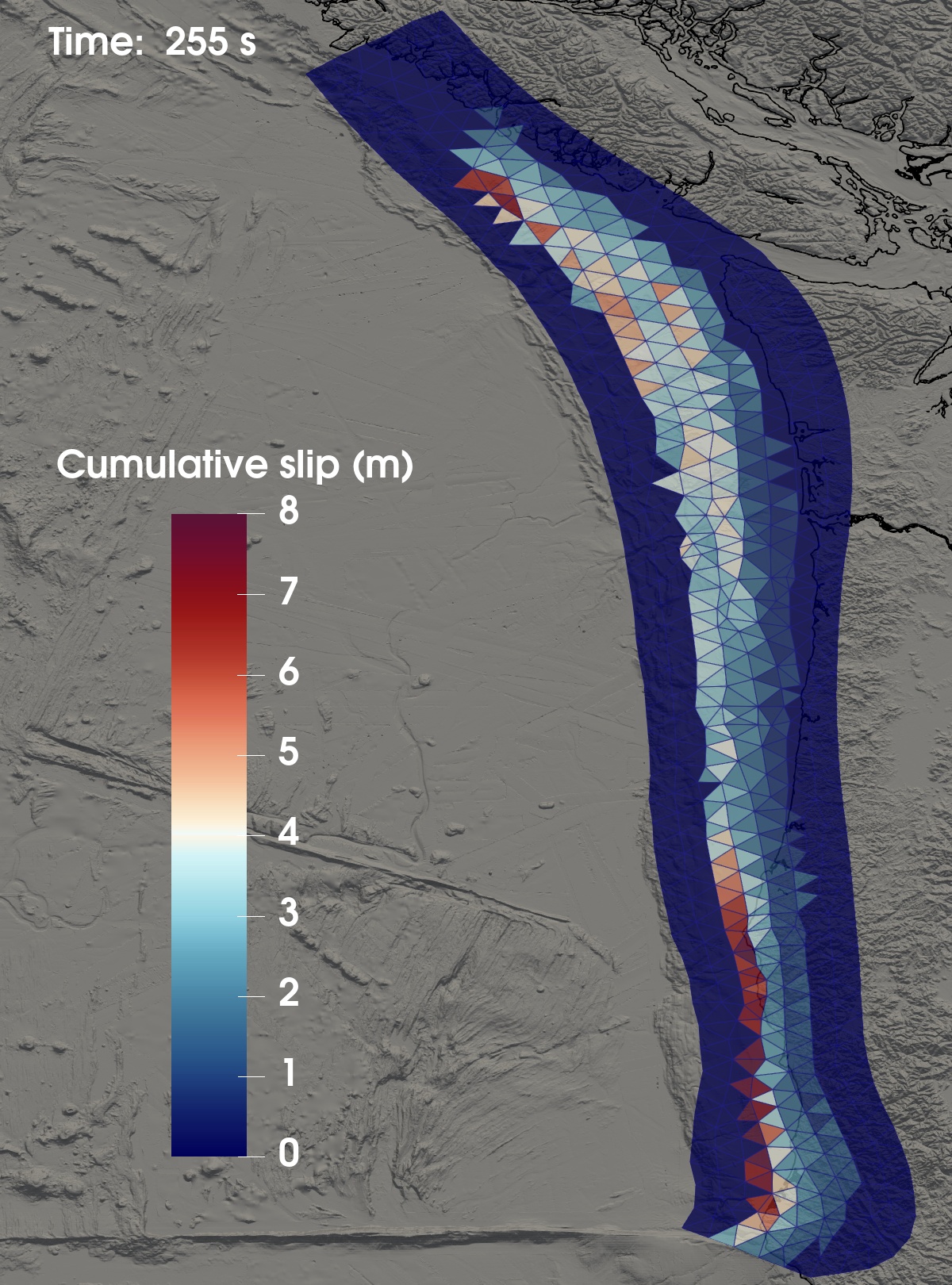}%
    \hfill%
    \includegraphics[width=0.33\textwidth]
    {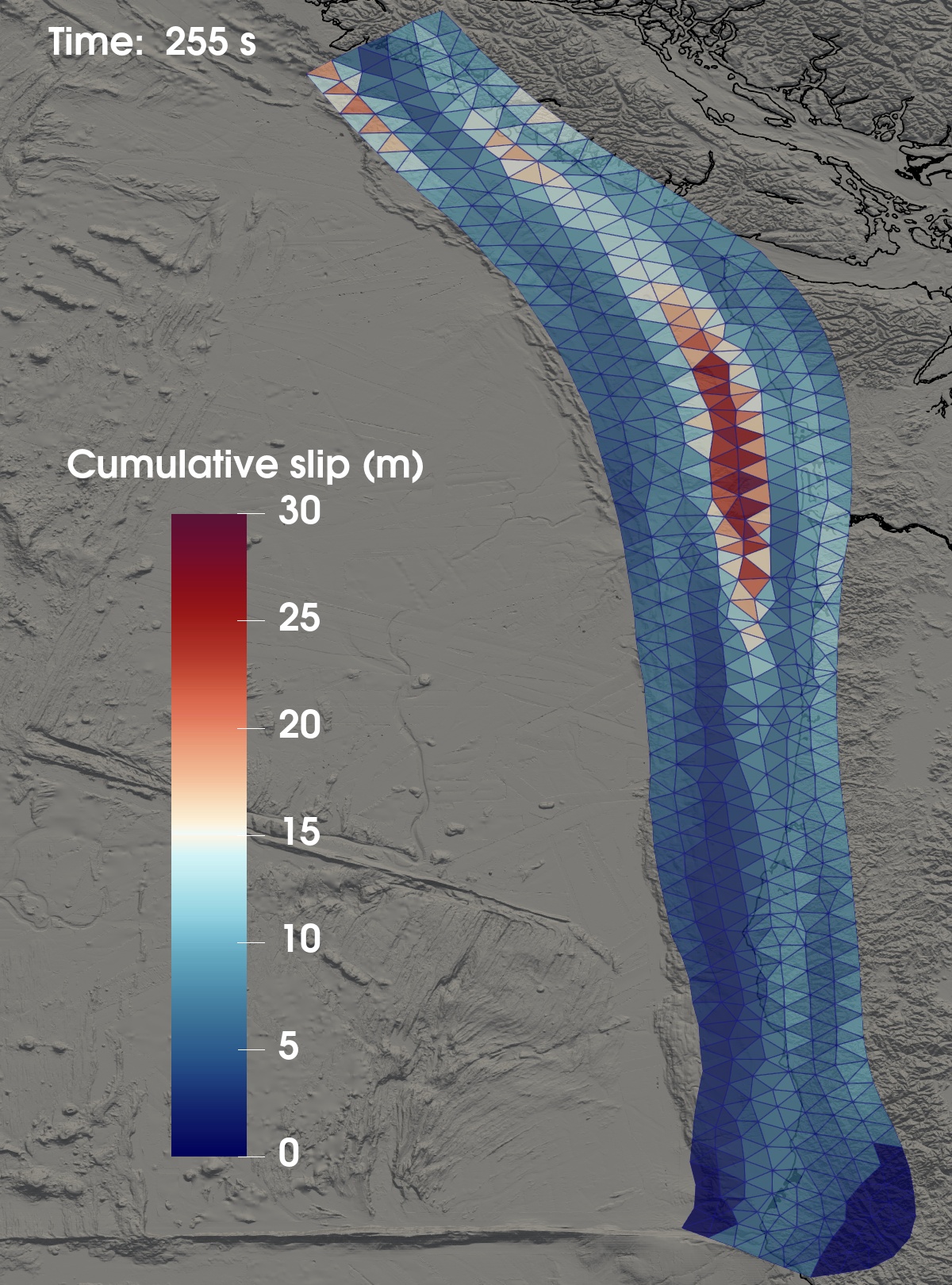}%
    \vspace{2pt}
    \includegraphics[width=0.33\textwidth]
    {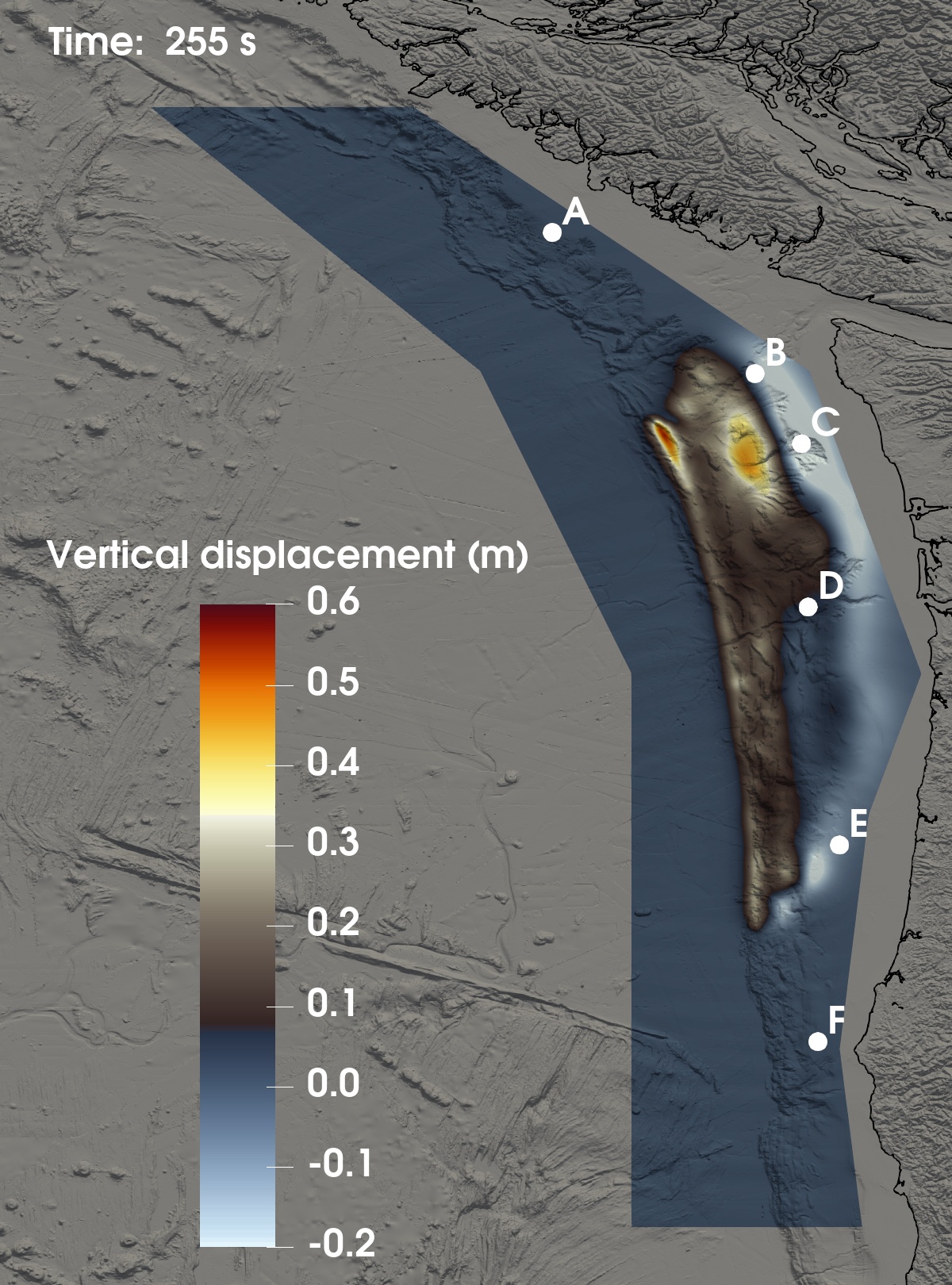}%
    \hfill%
    \includegraphics[width=0.33\textwidth]
    {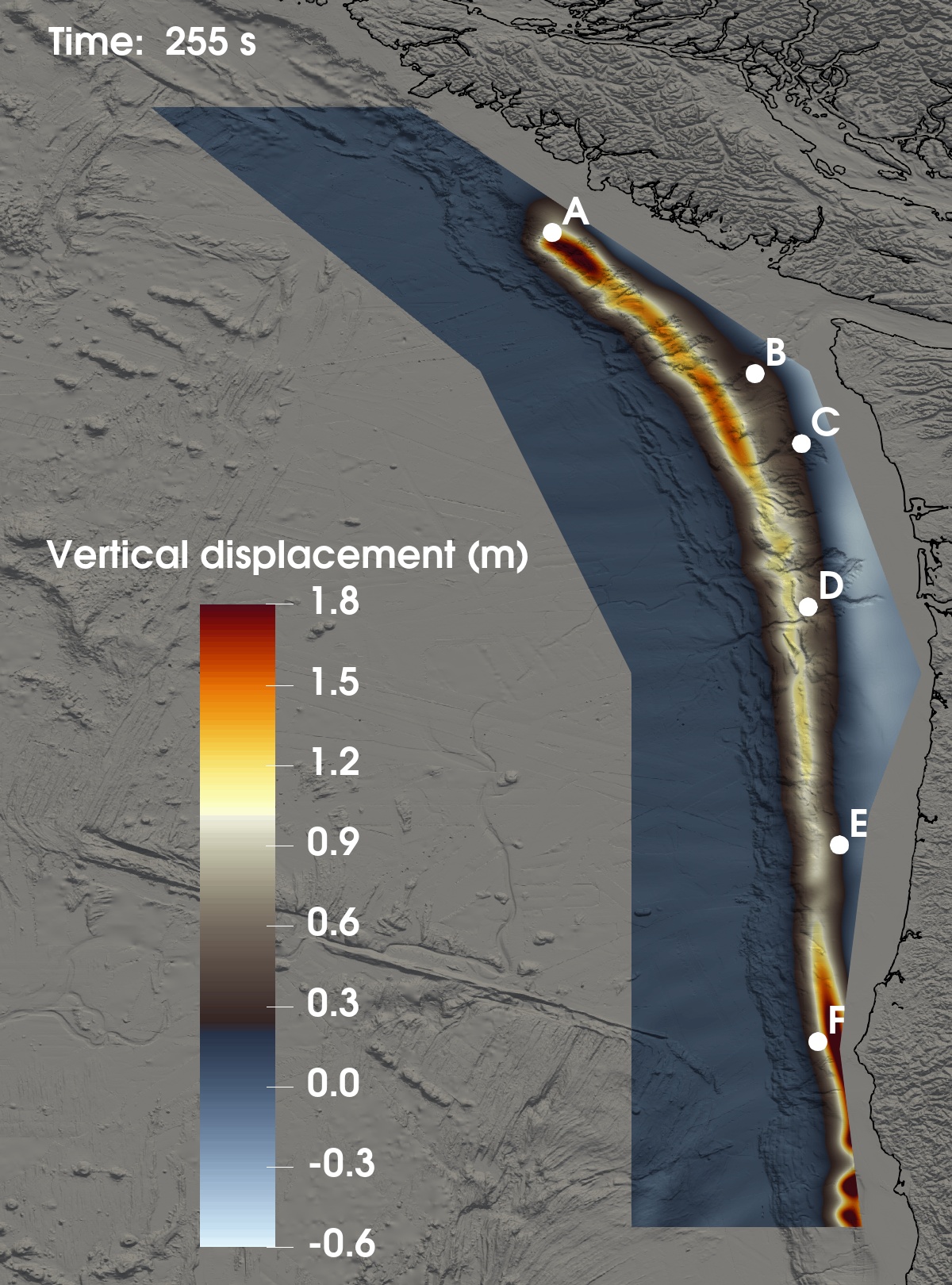}%
    \hfill%
    \includegraphics[width=0.33\textwidth]
    {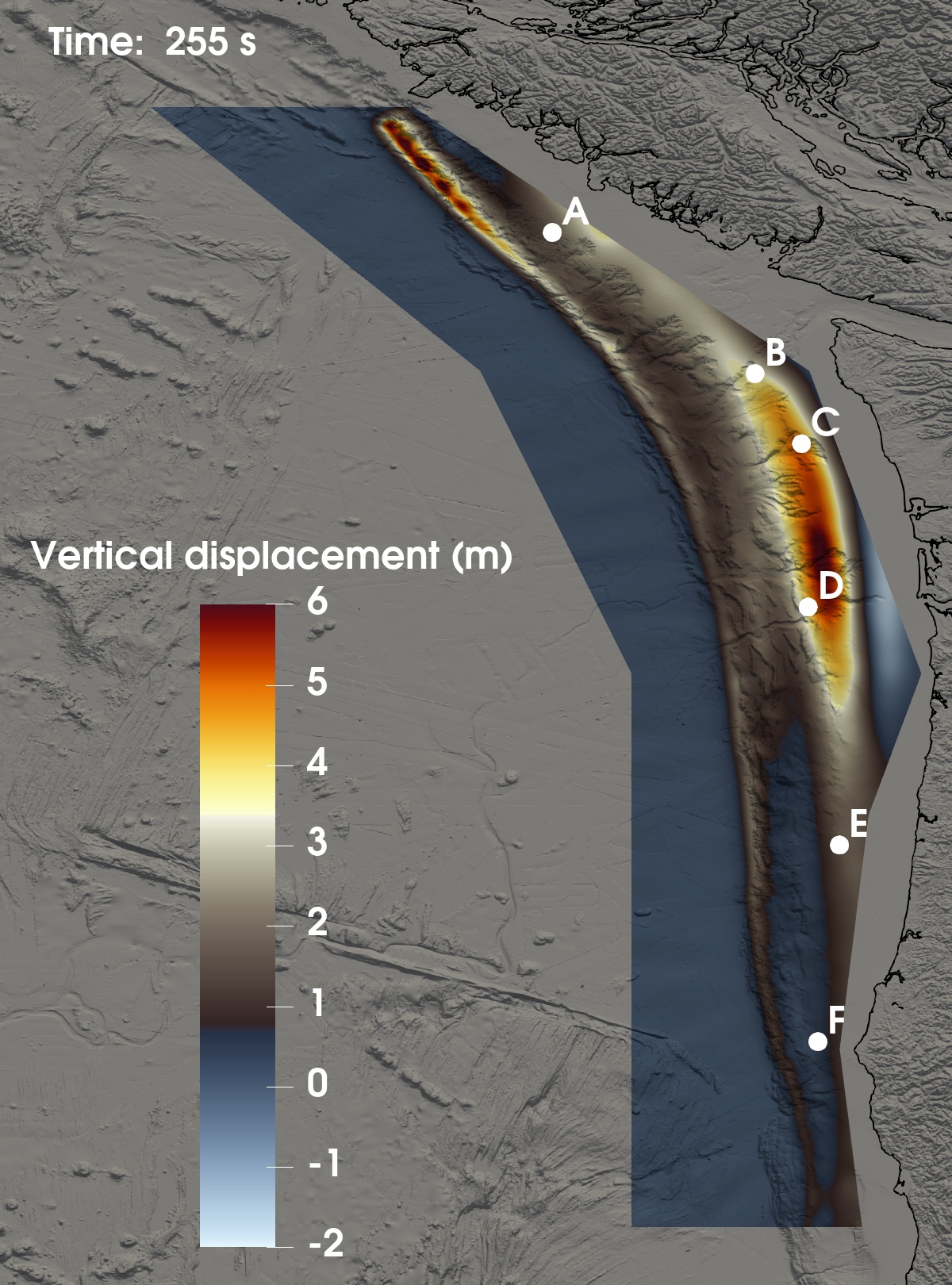}\\
    \vspace{2pt}
    \includegraphics[width=0.33\textwidth]{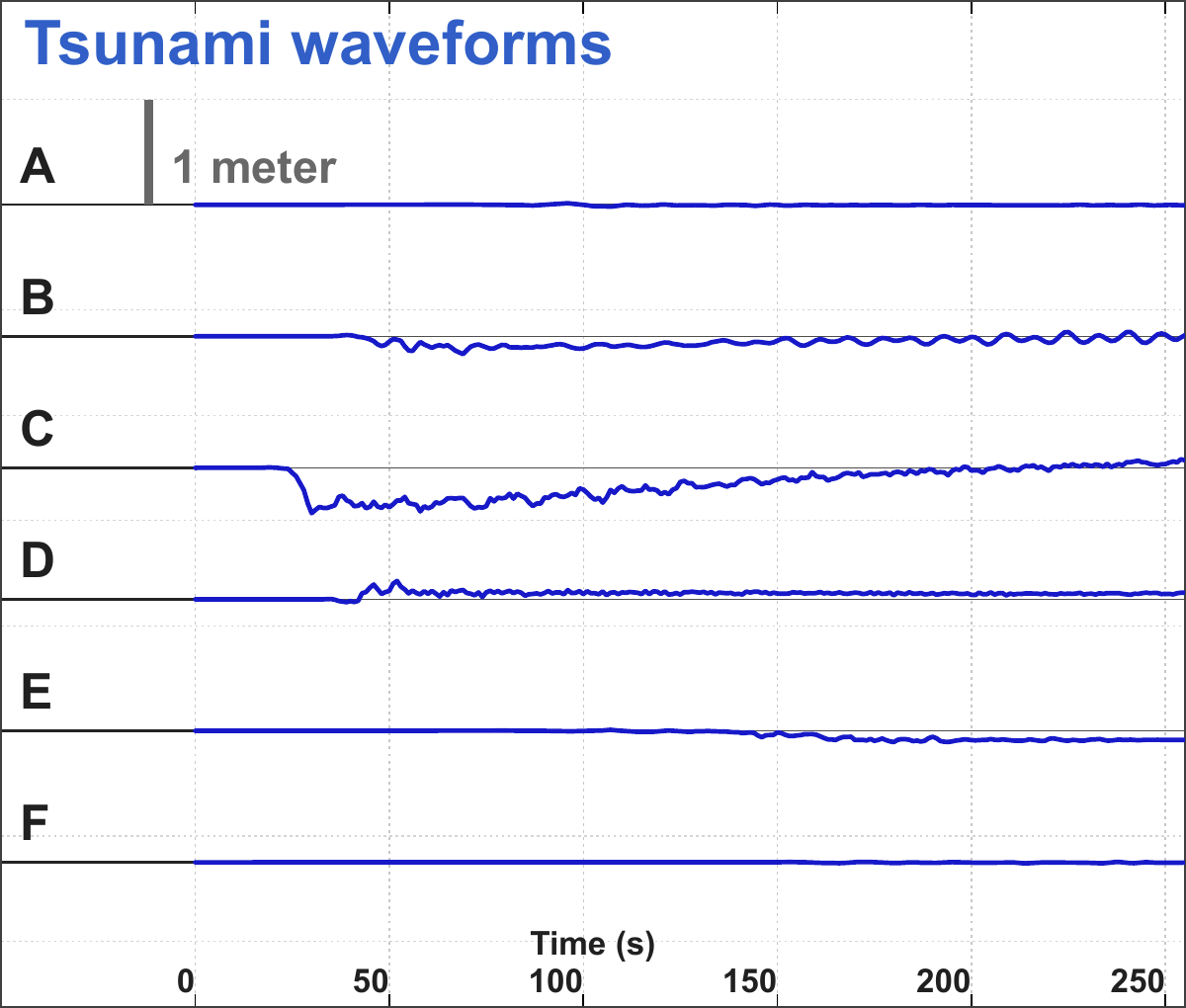}%
    \hfill%
    \includegraphics[width=0.33\textwidth]{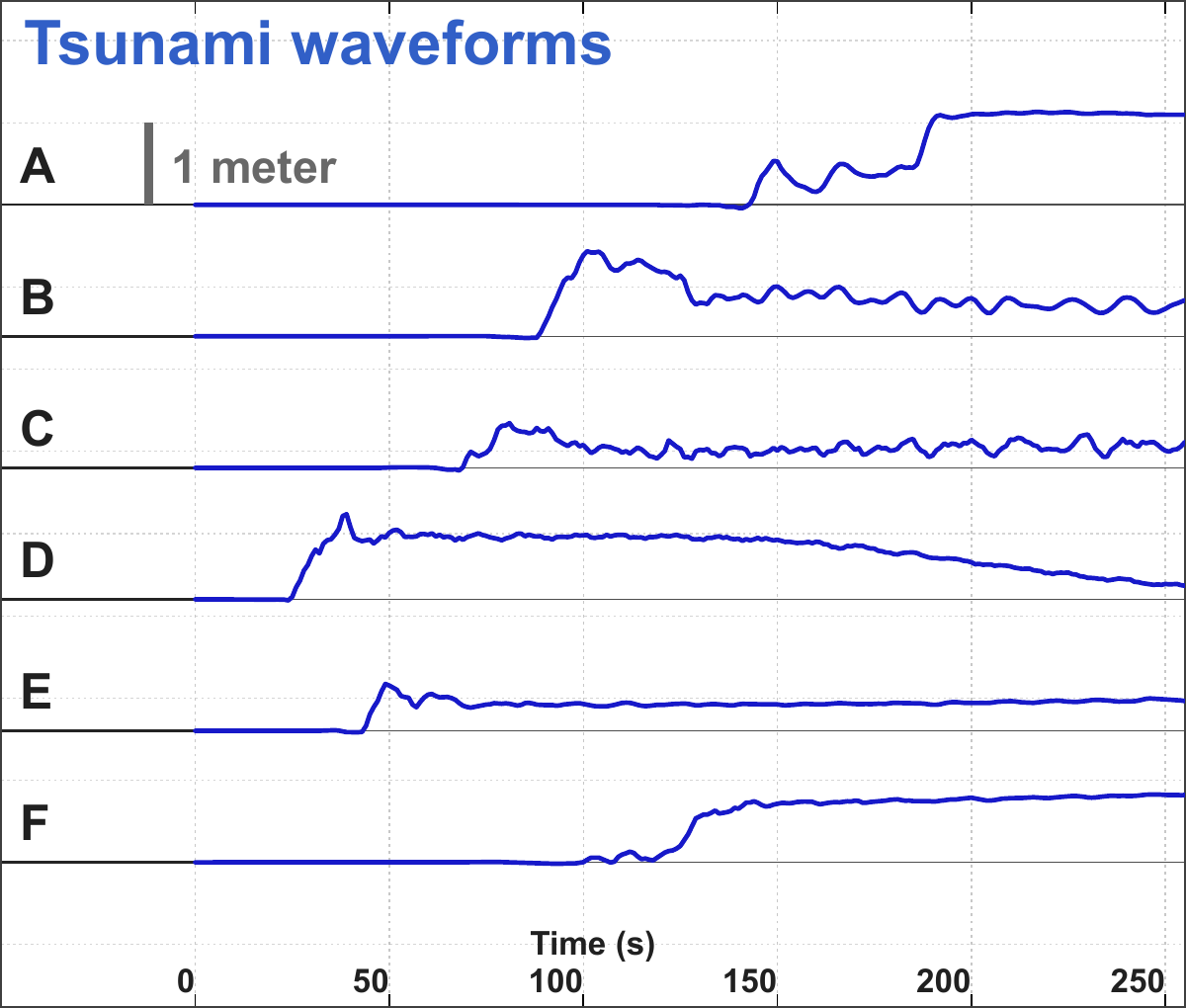}%
    \hfill%
    \includegraphics[width=0.33\textwidth]{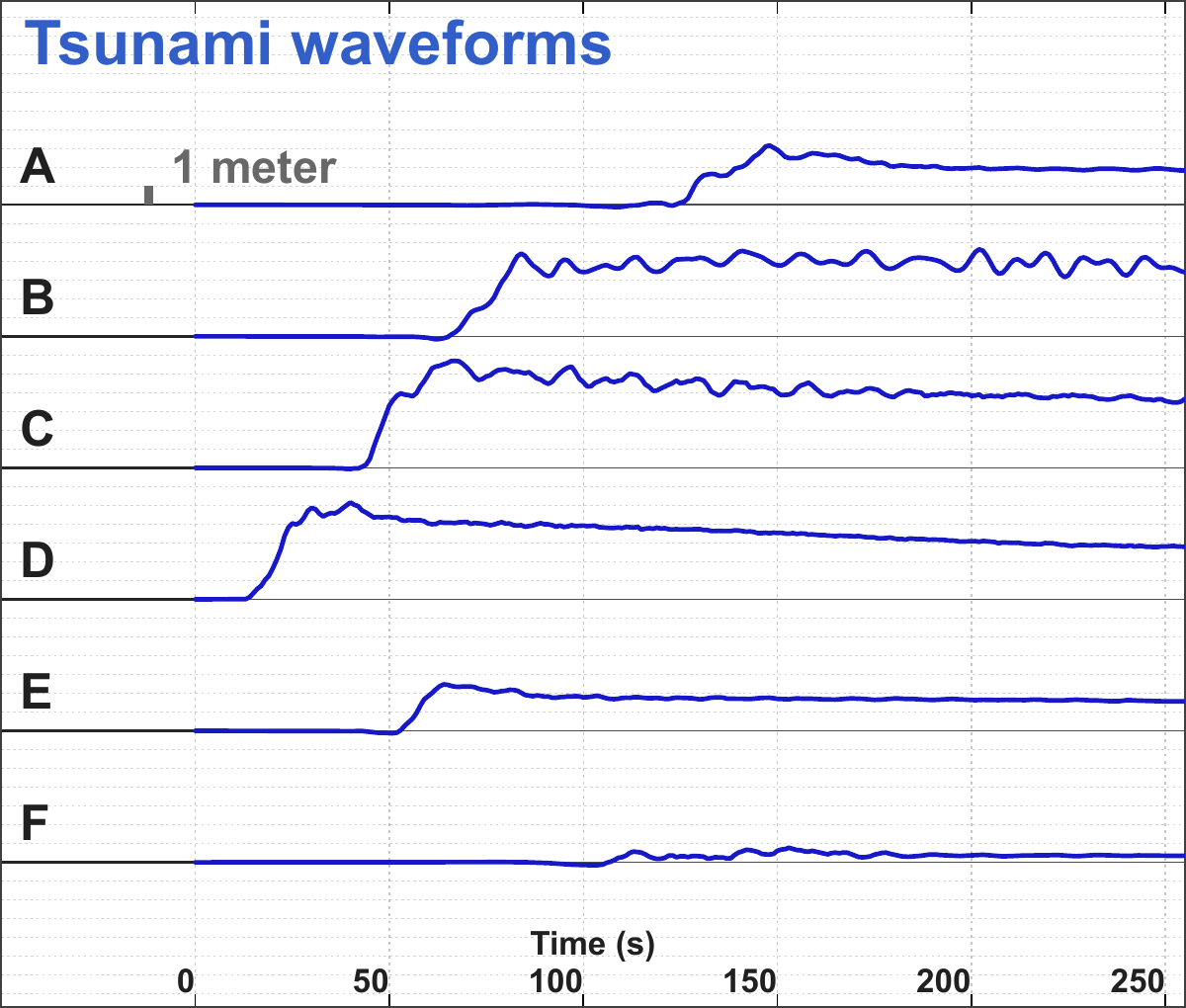}
    \caption{Snapshots of the cumulative slip distributions (top row) and vertical seafloor displacements (middle row) at \(t=255\,\mathrm{s}\), and the tsunami waveforms (bottom row) at a few sample locations (A--F, as indicated in the displacement plots), for three Cascadia FakeQuakes realizations with moment magnitude \(M_w=7.95\) (left column), 8.54 (middle column), and 9.09 (right column). Note that each panel uses an independent color range (slips and displacements) and axis scaling (waveforms). Hyperlinks to the following videos are provided: \href{https://youtu.be/ytFs_C-79jk}{\underline{cumulative slip animation}} and \href{https://youtu.be/eWfiEAEk9j0}{\underline{vertical displacement animation}}. The slip distributions represent the inputs to our earthquake-to-tsunami waveform generation pipeline, the seafloor displacements illustrate the (intermediate) outputs of the first convolution, and the tsunami waveforms are the (final) outputs obtained by the second convolution. Using 64 NVIDIA GB200 GPUs on TACC's Vista NVL72 system, the full pipeline applied to the Cascadia Subduction Zone executes in 24~ms per rupture.}
    \label{fig:paired-ruptures}
\end{figure*}

Figure~\ref{fig:paired-ruptures} shows the cumulative slip distributions on the Cascadia fault plane and
final-time seafloor displacement fields from the first convolution for three FakeQuakes realizations. The seafloor displacement fields, each with a spatiotemporal dimension of \(618{,}631{,}680\), are partitioned across GPUs and passed directly to the second convolution without being saved.

At this scale, the number of GPUs is determined by the required aggregate HBM capacity. 
The frequency-domain coefficients occupy 148.51~GiB per GPU; 
the measured resident state reaches 151.27~GiB per GPU, or
9.45~TiB across the distributed allocation. 
64 GH200 GPUs provide only 6.0~TiB, 
whereas 64 GB200 GPUs provide approximately 11.5~TiB of 
measured usable HBM within one NVL72 domain.

\begin{table}[htb]
\caption{Experimental configurations and time-to-solution of the\\ 
full pipeline applied to the Cascadia Subduction Zone}
\label{tab:full}
\centering
\footnotesize
\begin{tabular}{@{}lr@{}}
\toprule
Parameter & Value \\
\midrule
Number of subfaults $N_f$  & 963  \\
Number of waveform locations $N_q$ & 64 \\
Seafloor spatial grid points $N_m$& 2,416,530 \\
Timesteps $N_t$ & 256 \\
Machine & GB200 NVL72 domain \\
GPUs used & 64 of 72 \\
Persistent coefficients & 148.51~GiB/GPU \\
Resident state & 151.27~GiB/GPU\\ 
&(9.45~TiB aggregate) \\
\midrule
FFTMatvec time-to-solution & 23.998~ms/rupture \\
PyTorch time-to-solution & 522.205~ms/rupture \\
\bottomrule
\end{tabular}
\end{table}

Table~\ref{tab:full} reports the end-to-end (time-to-solution) performance 
after the initial one-time setup of loading the operators. 
FFTMatvec reduces end-to-end time per rupture by \(21.8\times\) relative
to the direct convolution approach and sustains 41.67 ruptures/s. At this measured
rate, applying the operators to an ensemble of \(10^4\) ruptures requires
approximately 4 minutes, compared with 87 minutes for the PyTorch
baseline implementation. 

\section{Conclusion}

We presented a multi-GPU pipeline for rapid earthquake-to-tsunami waveform generation that exploits the linear time-invariant structure of the underlying elastodynamic and acoustic--gravity wave equation models. Precomputed elastic Green's functions and acoustic--gravity adjoint responses reduce the source-to-waveform map to two consecutive convolution operators each represented by block lower-triangular Toeplitz matvecs. Our distributed FFTMatvec implementation evaluates these operators through blockwise Fourier transforms and batched frequency-domain matvecs, while retaining all intermediate quantities on-device and distributed across GPUs.

At full Cascadia scale, the pipeline maps 963 subfault slip-rate functions through 2,416,530 seafloor points to 64 waveform outputs over 256 timesteps. On 64 GB200 GPUs within one NVL72 domain, our pipeline evaluates the complete source-to-waveform map in 24~ms per rupture, a \(21.8\times\) speedup over the direct time-domain reference implementation using PyTorch. These results demonstrate our pipeline provides a practical route for generating large-scale waveform ensembles based on high-fidelity models for data-driven tsunami modeling and forecasting.

\section*{Acknowledgments}
This research was supported by DARPA COMPASS grant HR0011-25-3-0242, DOD MURI grant FA9550-24-1-0327, and DOE ASCR grant DE-SC0023171.

This research used resources from the National Energy Research Scientific Computing Center (NERSC) under allocations ALCC-ERCAP0030671, ScienceAtScale DDR-ERCAP0034808 and NESAP DDR-ERCAP0038013. 
The authors acknowledge the Texas Advanced Computing Center (TACC) at The University of Texas at Austin for providing computational resources that have contributed to the research results reported within this paper.

\bibliographystyle{IEEEtran}
\bibliography{references}

\end{document}